\documentclass[aip,reprint,showpacs,prb,amssymb,amsmath,floatfix,longbibliography]{revtex4}

\usepackage{graphicx}
\usepackage{dcolumn}
\usepackage{bm}
\usepackage{color}
\usepackage{underscore}
\begin{document}

\title{Thermoelectric properties of $n$-type SrTiO$_{3}$}

\author{Jifeng Sun}
\author{David J. Singh}
\email{singhdj@missouri.edu}

\affiliation{Department of Physics and Astronomy, University of Missouri,
Columbia, Missouri 65211-7010, USA}

\date{\today}

\begin{abstract}
We present an investigation of the thermoelectric properties of cubic perovskite SrTiO$_{3}$.  The results are derived from a combination of calculated transport functions obtained from Boltzmann transport theory in the constant scattering time approximation based on on the electronic structure and existing experimental data for La-doped SrTiO$_{3}$. The figure of merit $ZT$ is modeled with respect to carrier concentration and temperature. The model predicts a relatively high $ZT$ at optimized doping, and suggests that the $ZT$ value can reach 0.7 at $T$ = 1400 K. Thus $ZT$ can be improved from the current experimental values by carrier concentration optimization.
\end{abstract}

\maketitle

\section{\label{sec:level1}INTRODUCTION}

Themoelectric (TE) energy conversion has been widely recognized as a promising technology in power generation including waste heat recovery \cite{r0,r1,r2}. However the applications are limited due to the low efficiency. This is associated with the dimensionless figure of merit $ZT = \sigma S^{2}T/\kappa$, where $\sigma$ is the electrical conductivity, $S$ is the Seebeck coefficient, $T$ is the temperature and $\kappa$ is the thermal conductivity. 
Oxides may offer advantages in applications.
\cite{matsubara,tsubota,he,saucke}
However, while there are a number of thermoelectric materials with $ZT>$1 \cite{r3,r4,r5}, progress in oxide thermoelectrics has been slower.

Oxide thermoelectrics with high $ZT$ value approaching 1 mainly fall in $p$-type Co oxides and related alloys \cite{r6,r7,r8,r9}.  $n$-type oxide thermoelectrics with high performance are rare and with $ZT$ value barely exceeding 0.5. The best $n$-type to date is found in ZnO ceramics at high temperatures \cite{r10,r11,r12,r13}. Among the several $n$-type oxides, SrTiO$_{3}$ (STO) with cubic perovskite structure has attracted growing attention for TE power generation at high temperatures due to its large Seebeck coefficient originated from the
degenerate Ti 3$d$-t$_{2g}$ band at the conduction band minimum and high power factor (PF) comparable to that of Bi-Te alloy \cite{r14}. However the $ZT$ value achieved is still quite low ($<$ 0.5) even at 1000 K. This is mainly attributed to its high thermal conductivity (6-12 Wm$^{-1}$K$^{-1}$ for undoped STO \cite{r15}). Optimizing the doping level and elements as well as introducing point defects are effective methods to reduce the thermal conductivity. For instance the La- and Nb-doped STO can have $\kappa$ as low as 3 Wm$^{-1}$K$^{-1}$ at about 1000 K depending on the doping concentrations \cite{r16}. Lower thermal conductivity down to 2.3 Wm$^{-1}$K$^{-1}$ at 1073 K can be achieved through double doping with La and Dy on the Sr (A) site \cite{r17}. The $ZT$ at corresponding temperature is reported to be 0.36 which is a relatively high value in doped STO. A-site vacancies also can lower $\kappa$. Particularly, Popuri $et$ $al$. has shown a glass-like thermal conductivity through A-site vacancies and $\kappa$ = 2.5 Wm$^{-1}$K$^{-1}$ can be realized at 1000 K \cite{r18}. Recently, Lu and co-workers find high $ZT$ about 0.4 at 973 K with similar low $\kappa$ = 2.5 Wm$^{-1}$K$^{-1}$ from their La-doped and A-site-deficient samples \cite{r19}. Importantly, STO has been identified as a material with a corrugated band structure of a form that is particularly beneficial for the electronic part of $ZT$, but which at the same time is not reasonably described by standard parabolic band models. This makes detailed calculations particularly important.     

Here we report on the thermoelectric properties of cubic STO from a combination
of first-principles calculations and analysis of existing experimental data.
High temperature up to 1400 K was investigated because the thermoelectric
performance typically increases with temperature for wide band gap materials.
Furthermore, the melting point of STO is high (2080 $^{\circ}$C)
and $n$-type STO is expected to be applicable at high temperatures.
We find that the optimized $n$-type $ZT$ can reach a peak value
of $\sim$ 0.7 at 1400 K with reasonable high doping concentrations. 

\section{\label{sec:level2}Computational methods}

The model for $n$-type STO was constructed along the lines of the prior model
for PbSe \cite{r20}.
Specifically, the transport functions and coefficients
are obtained using a relaxation time
approximation to Boltzmann transport theory \cite{ziman} based on
first-principles electronic structure.
We use the constant scattering time approximation (CSTA) also
known as the constant relaxation time approximation,
as implemented in the BoltzTraP code \cite{r21}.
The CSTA consists in taking the energy dependence of the band structure
as the main ingredient in the energy dependence of the conductivity
in the transport formula. \cite{r21,zhang}
With this approximation the scattering time cancels in the expression
for the thermopower, so that one can obtain the thermopower as a function
of temperature and doping level from the band structure without any
additional input.

We emphasize that while the CSTA neglects the energy dependence of the
scattering rate on energy scales of $k_BT$, as compared to the energy
dependence of the electronic structure, it does not involve any
assumption about the temperature or doping level dependence of the
scattering.
Furthermore, the CSTA has been used to successfully describe the
Seebeck coefficients and their temperature and doping dependencies
in a wide variety of thermoelectric materials. This includes both
conventional thermoelectrics \cite{r20,r21,yang,r22,fei,bjerg},
and even materials where unusual scattering may be expected such as
the oxide Na$_x$CoO$_2$, \cite{singh}
and PdCoO$_2$, \cite{ong}
which was recently shown to display hydrodynamic electron transport.
\cite{moll}
We note that it has been shown to yield results in accord with experiment
for SrTiO$_3$, \cite{kinaci} which is the subject of the
present study.
In addition the CSTA
has proven useful as a basis for performing high throughput
searches for new thermoelectrics. \cite{madsen,wang,curtarolo}
Going beyond the CSTA would require detailed knowledge of the energy
dependent scattering mechanisms including different sources of scattering
in combination. For example, if one includes only acoustic phonon
scattering with $\tau^{-1}$ proportional to the density of states,
one obtains the result that the mobility diverges and the conductivity
is flat as the carrier concentration is lowered to zero.
The behavior can be understood from the low $T$ behavior,
specifically the density of states is given
by $N(\epsilon)=\partial n(\epsilon)/\partial\epsilon$, which for
a power law dependence of $n$ on $\epsilon$ yields
$N(\epsilon)\propto n(\epsilon)/\epsilon$. $\sigma=ne\mu$,
where $\mu$ is the mobility. For a parabolic band
$\epsilon=\hbar^2k^2/2m$,
$n(\epsilon)\propto\epsilon^{3/2}$,
$\sigma\propto e^2\tau n/m$,
while density of states $N(\epsilon)$ goes as $\epsilon^{1/2}$, i.e.
$n^{1/3}$ so that $\sigma$ has a sublinear dependence on $n$, with a 
divergent mobility, $\mu=\sigma/ne$. At finite $T$, the flatness of
$\sigma/\tau$ is enhanced by the Fermi broadening and the divergence of
$\mu$ becomes stronger.

This unphysical result comes from ignoring point defect, polar
optical phonon and other scattering. It is 
the case for a parabolic band model and
we verified by direct calculation at 300 K
that it is also so for the band structure of SrTiO$_3$.
Defect scattering, will in general limit the mobility and prevent such
divergences of the mobility.
SrTiO$_3$ has a large dielectric constant due high Born charges leading
to nearness to ferroelectricity. This provides a mechanism for screening
of ionized impurity and other defect scattering. \cite{du}
In SrTiO$_3$ the dielectric constant and therefore this source of
screening is a strong function of temperature.
The polar nature of the compound and the high Born charges lead to the
expectation of strong polar optical phonon scattering.
Himmetoglu and co-workers find that at room temperature with moderate
carrier concentration this scattering mechanism dominates the
mobility and that it has a strong and complex momentum
dependence, \cite{himmetoglu}
and a complex energy dependence may also be expected in analogy
with other materials. \cite{komirenko}
Furthermore, one expects acoustic phonon scattering, discussed above,
to contribute to scattering in any material, and electron-electron
scattering may also contribute for heavy doping and low temperature.

It is possible to fit various energy, doping and temperature dependent
scattering terms to experiment for
semiconductors for which large amounts of doping and temperature
dependent data is available. An example is provided by the detailed
fits performed for the Pb chalcogenide IV-VI semiconductor thermoelectrics,
which reproduced extremely well the known body of experimental data.
\cite{ravich1,ravich2,ravich3}
These models may, however, have limited predictive ability outside the range
of known data due to the large number of fitting parameters used.
For example, they lead to the expectation that the $ZT$ for
PbTe could not exceed unity and that PbSe would invariably be inferior
to PbTe. In contrast, calculations based on the CSTA, similar to the
approach followed here, did predict that ordinarily (Na and/or K)
but heavily doped p-type PbTe and PbSe would
both have high $ZT$ above unity,
\cite{r20,singh-pbte} as was subsequently shown to be the case. \cite{pei}
Therefore we use the CSTA, which while approximate,
allows one to predict the thermopower
directly from the first principles electronic structure, with no
adjustable parameters.

In the present work, we use results calculated from the
electronic structure using the CSTA and combine these
models derived by fitting experimental data
for the temperature and doping concentration
dependent relaxation time $\tau$ and lattice thermal conductivity,
$\kappa_{l}$, to obtain the conductivity and thermal conductivity.
Eventually we are able to get the figure of merit $ZT$ in terms of $T$ and $n$. 

The band structure calculations
were performed with the linearized augmented plane wave (LAPW) method
\cite{r25}, as implemented in the WIEN2K code \cite{r26}.
We used LAPW sphere radii of 2.5 bohr for Sr, 1.9 bohr for Ti and 1.75 bohr for O. A k-point mesh of 6$\times$6$\times$6 was used for total energy calculations and much denser k meshes are used for density of states (17$\times$17$\times$17), isoenergy surfaces and transport properties (27$\times$27$\times$27). We adopted the experimental cubic structure with lattice constant a = 3.906 \AA\,\cite{r27}, with the modified Becke-Johnson (mBJ) potential of Tran and Blaha \cite{r28}. This potential generally gives improved band gaps for simple semiconductors and insulators \cite{r20,r28,r29}. Spin-orbit coupling (SOC) was included in all the calculations. 

\section{\label{sec:level3}RESULTS AND DISCUSSION}

\begin{figure}
\includegraphics[width=\columnwidth]{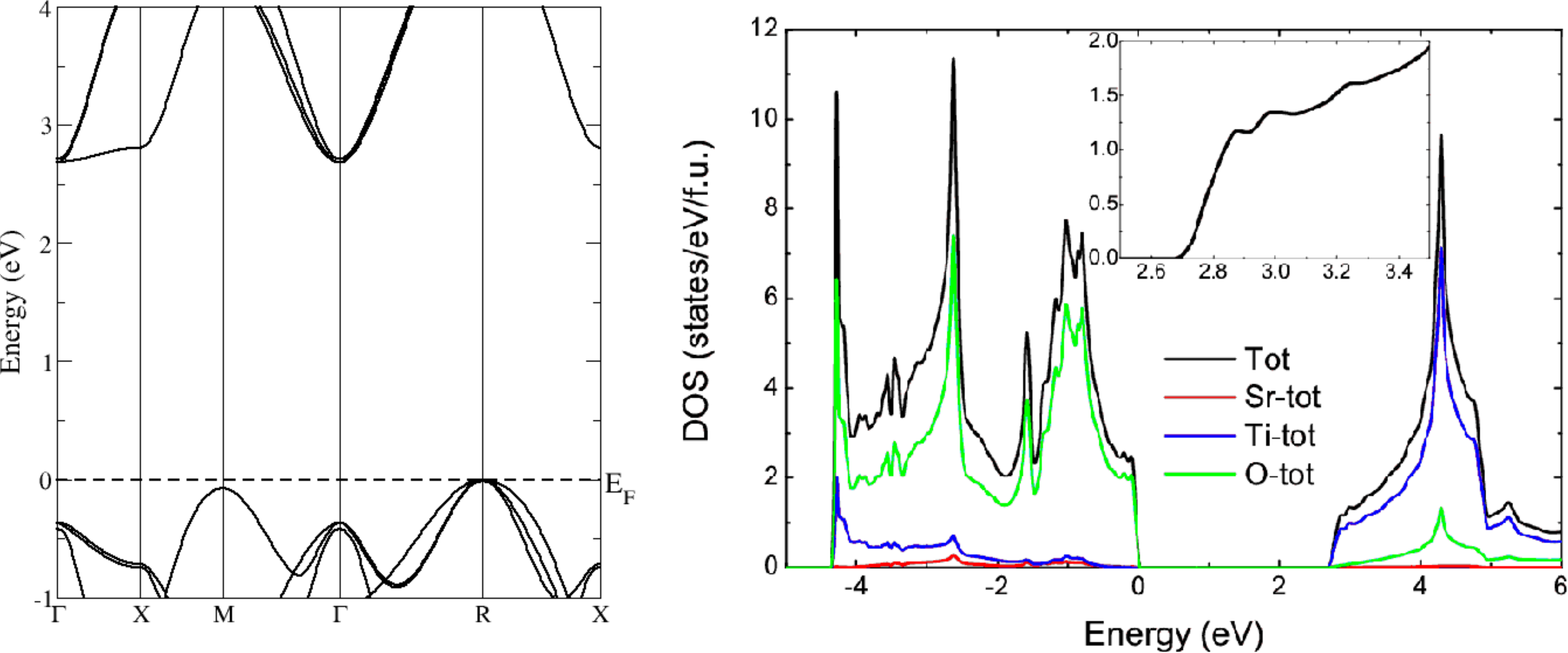}
\caption{\label{fig1} Calculated band structure and total density of states of SrTiO$_{3}$ with mBJ and spin-orbit coupling. The inset shows the detailed feature near CBM. Density of states are shown per SrTiO$_{3}$ formula unit (f.u.).  Energy zero is set at the valence band maximum.}
\end{figure}

SrTiO$_{3}$ is a band insulator with a band gap value of 3.2 eV.
\cite{cardona}
Our calculated band structure agrees with this quite well
as shown in Fig. \ref{fig1}. The calculated band gap is about 2.68 eV (R-$\Gamma$) with SOC using mBJ potential. At the $\Gamma$ point near the conduction band minimum (CBM), the lowest three bands are largely from Ti 3$d$-t$_{2g}$ states and the degeneracy is lifted by SOC effects with a split-off energy of about 30 meV. This agrees reasonably with other theoretical calculations (25-28 meV \cite{r30,r31}) and experiment (17 meV \cite{r32}) by interband Raman scattering. The electronic density of states (DOS) is presented in the right panel of Fig. \ref{fig1}, where the inset shows the details at the CBM. As known, the valence band is mainly derived from oxygen 2$p$ states and the conduction band is dominated by Ti 3$d$ states especially near CBM. In fact the high $S$ in STO is mainly due to the large DOS effective mass \cite{r33,r34}. 

Within a simple parabolic band model, the $S(T)$ is proportional to effective mass ($m^{\ast}$) and decreases as the 2/3 power of doping concentration.
Moreover, $m^{\ast}$ plays different roles in $\sigma$ and $S$. For thermoelectric performance it is important to have both high $S$ and high $\sigma$, which as discussed previously \cite{r41} have opposite dependences on both carrier concentration and effective mass. This conundrum can be resolved by certain complex band structures that exploit the different transport integrals that enter $\sigma$ and $S$ \cite{r42,r43,r44,r45,r46,r47,r48}. STO has a particular band shape arising from the degeneracy of the $t_{2g}$ levels in an octahedral crystal field that gives effectively lower dimensional behaviour in transport even though the material is cubic, analogous to the case of PbTe \cite{r47}. The sharp onset of the DOS at the CBM reflects 
the low-dimensional nature of the electronic structure.
  
\begin{figure}
\includegraphics[width=\columnwidth]{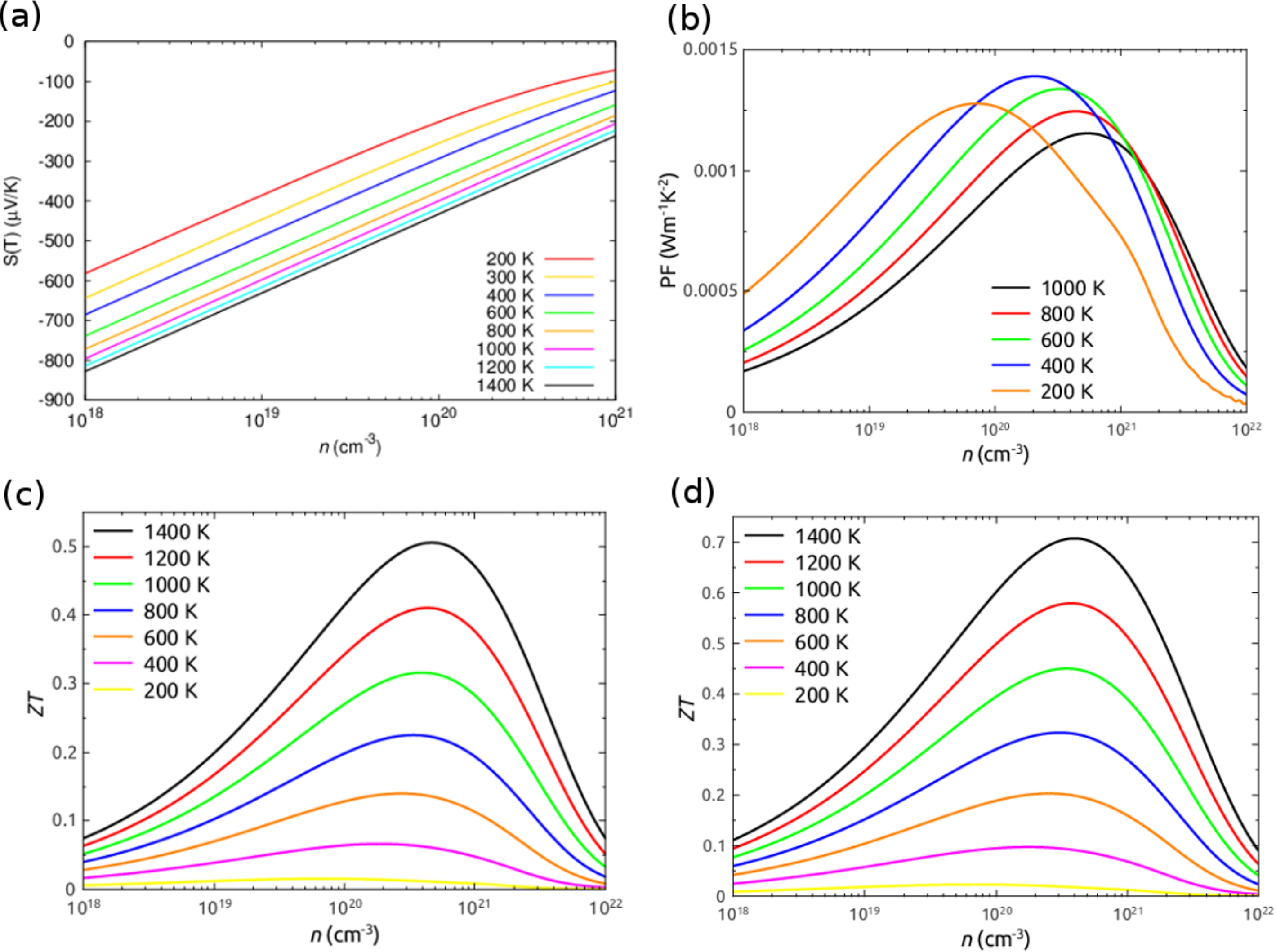}
\caption{\label{fig2} Calculated thermoelectric properties for $n$-type STO: (a) $S(T,n)$, (b) power factor , (c) $ZT$ using the bulk $\kappa$ at 300 K and (d) $ZT$ using smaller $\kappa$ (see text). }
\end{figure}

The temperature and doping concentration dependent Seebeck coefficient, $S(T,n)$, can be obtained directly from the electronic structure with no adjustable parameters within CSTA. From the Wiedemann-Franz relation one can rewrite $ZT$ as $ZT = rS^{2}/L$ where $r = \kappa_{e}/(\kappa_{e}+\kappa_{l})$ and $L = 2.45\times10^{-8}$ W$\Omega/$K$^{2}$ is the standard Lorenz number. If $\kappa_{l}$ is neglected assuming the upper bound of $ZT$ which can reach unity when $S$ = 156 $\mu$V/K. In reality good TE materials usually have $S$ larger than 200 $\mu$V/K. We illustrate the calculated $S(T,n)$ of $n$-type STO at different temperatures in Fig. \ref{fig2} (a). As seen, high Seebeck coefficients ($>$ 200 $\mu$V/K ) are obtained even at quite low temperatures (200 K) with doping concentration up to 10$^{20}$ cm$^{-3}$. At high temperatures ($>$ 1000 K), $S$ exceeds 200 $\mu$V/K at high doping concentrations ($\sim$ 10$^{21}$ cm$^{-3}$). Importantly, due to the substantial band gap, there is no bipolar effect and therefore $S$ keeps increasing with $T$ to high temperatures even at low doping concentrations.

Within the framework of CSTA, one can obtain $\sigma/\tau$ directly from the electronic structure as a function of $n$ and $T$.
However, it is not possible to get $\sigma$ without the knowledge of the scattering rate $\tau^{-1}$.
Therefore, in order to proceed, we used the strategy of our recent work on Mg$_{2}$(Ge,Sn) \cite{r35} and a prior model on PbSe where the experimental data was used to fit the relaxation time $\tau$. For this purpose, the extrapolation of the experimental data was based on the recent work of La-doped STO films grown by hybrid molecular beam epitaxy, conducted by Cain and co-workers \cite{r36}. Their samples have relatively higher electrical conductivities at the same conditions ($T$ and $n$) compared to other studies, which may indicate the higher quality of the STO films. Specifically, we used the data set which has the highest $\sigma$ and choose the experimental $S$ at corresponding $T$ with a value of about -270 $\mu$V/K at 300 K. The reason is that we want to estimate the optimum $ZT$, and the choice of the best samples as the baseline minimizes the effects on the predicted $ZT$ due to extrinsic effects that are presumably controllable. By comparing with our calculated $S(T,n)$ (Fig. \ref{fig2} (a)), we obtain the carrier concentration $n = 8.4\times10^{19}$ cm$^{-3}$. This calculated $n$ is smaller than the reported one (2$\times10^{20}$ cm$^{-3}$). We are using the experimental $S$ rather than the the nominal carrier concentration because of the potential neglect of compensating defects as well as the fact that the carrier concentration is not the same as in the Hall measurements for a non-parabolic case. At this temperature and doping concentration, combing with the calculated $\sigma/\tau$ and experimental $\sigma$ ($\sim$ 18500 $\Omega^{-1}$m$^{-1}$)\cite{r36} yields $\tau$ = 7.2$\times$10$^{-15}$ s. In this regime, we considered an approximate electron-phonon $T$ dependence of $\tau$ where $\tau$ is proportional to $T^{-1}$ and decreases with carrier concentration as $n^{-1/3}$. Therefore these together with fitted $\tau$ yield $\tau$ = 9.45$\times10^{-6}T^{-1}n^{-1/3}$ with $\tau$ in s, $T$ in K and $n$ in cm$^{-3}$. Thus $\sigma$ can be calculated through $\sigma/\tau\times\tau$.
The corresponding PF ($\sigma S^{2}$) obtained from the calculated
$S(T,n)$ and $\sigma$ as mentioned above is presented in Fig. \ref{fig2} (b).
The maximum PF is about 1.4$\times$10$^{-3}$ Wm$^{-1}$K$^{-2}$ at 400 K,
which is smaller than the heavily La doped STO
(3.6$\times$10$^{-3}$ Wm$^{-1}$K$^{-2}$) at room temperature \cite{r14}. 
Moreover, the position of the peak in the PF as a function of carrier
concentration
shifts to higher doping levels with increasing $T$.      
This means that optimization of the carrier concentration
for high temperature will lead to higher doping levels than are optimum at
lower $T$.
Thermal conductivity is also required to assess $ZT$.
As mentioned, $\kappa$ is composed of both the lattice and electronic contributions. The electronic part can be directly calculated using the Wiedemann-Franz relation from $\kappa_{e} = L\sigma T$. The lattice thermal conductivity usually behaves as $T^{-1}$ until high temperatures, as also observed in the experiments \cite{r16}. Hence we estimate $\kappa$ as $\kappa = A/T+L\sigma T$ with constant A determined by fitting the experimental data of $\kappa_{l}$. There have been a number of studies of the thermal conductivity measurements on both the bulk and doped STO at elevated temperatures \cite{r15,r16,r17,r18,r19,r37,r38,r39,r40}. Due to the significant difference of $\kappa$ between bulk and heavily doped materials, we consider two scenarios with one taking the bulk $\kappa$ at 300 K ($\sim$ 11 Wm$^{-1}$K$^{-1}$)\cite{r16,r39} and the other using the doped $\kappa$ at the same $T$ ($\sim$ 8 Wm$^{-1}$K$^{-1}$, based on moderately doped samples \cite{r16,r19}). With the second scenario, we are able to get low $\kappa$ from 2.3 - 2.5 Wm$^{-1}$K$^{-1}$ at 1000 K. These values are in better accord with experiments at high temperatures  ($\sim$ 1000 K).

\begin{figure}
\includegraphics[width=\columnwidth]{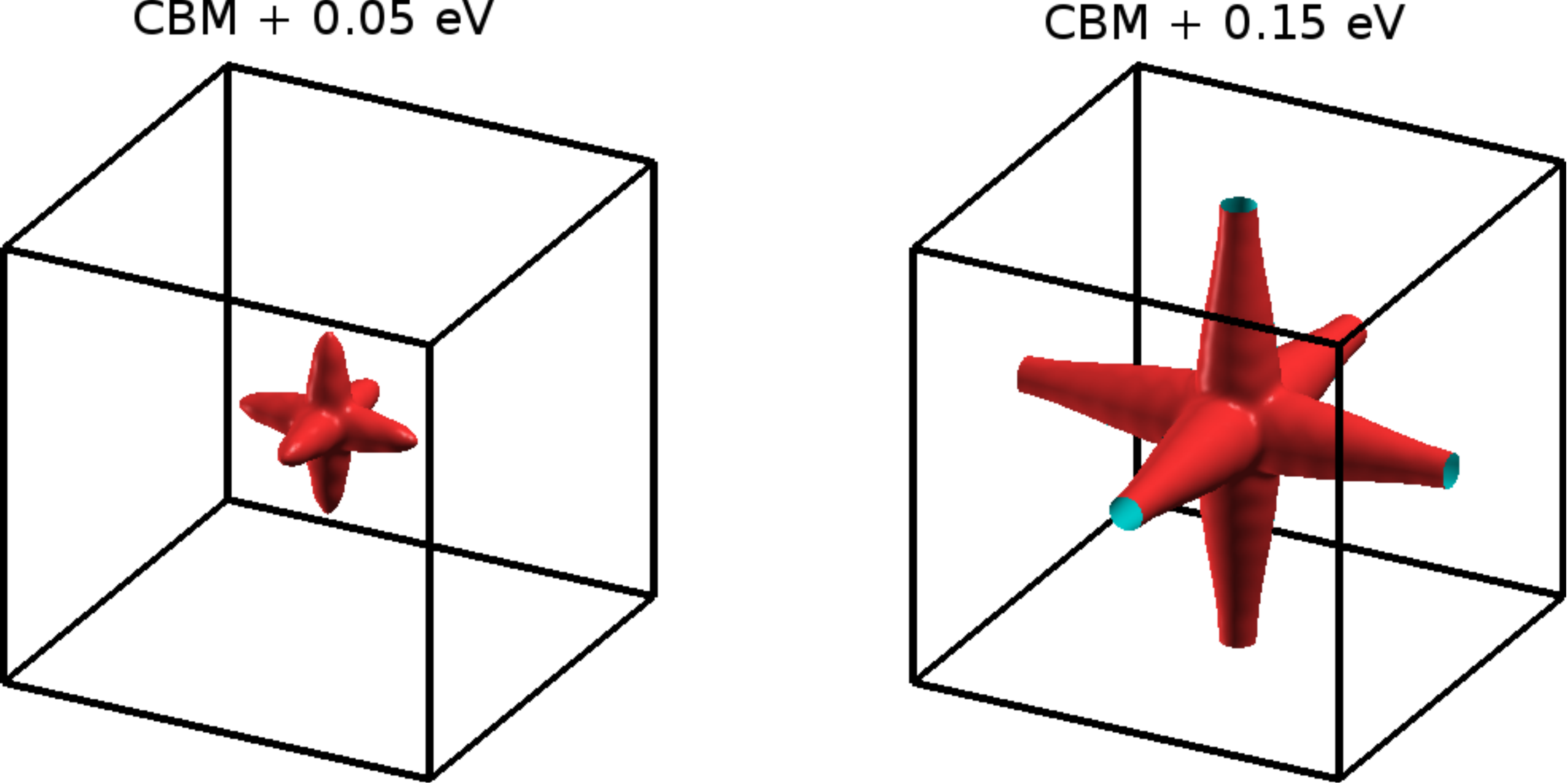}
\caption{\label{fig3} Isoenergy surfaces at 0.05 eV and 0.15 eV above CBM for $n$-type STO. Note the low-dimensional-like intersecting cylinders in higher carrier concentration (right panel) that originated from the $t_{2g}$ orbitals is beneficial for achieving high $ZT$. }
\end{figure}

Finally, we obtain the $ZT$ as a function of temperature and doping concentration. The results are shown in Figs. \ref{fig2} (c) and (d) with (c) depicting the data from model with bulk $\kappa$. Although it has a number of assumptions, the model can be used to describe the behaviour of $ZT$ with respect to $T$ and $n$. As seen, $ZT$ strongly depends on $T$ and increases up to the highest $T$. This is mainly due to the substantial band gap so that the Seebeck coefficient keeps increasing to high $T$. The lattice thermal conductivity decreases at high $T$ but the electronic part has weaker $T$ dependence. The maximum $ZT$ with bulk $\kappa$ (Fig. \ref{fig2} (c)) is about 0.5 at 1400 K and 0.3 at 1000 K with high doping concentration range ($\sim$4-5$\times10^{20}$ cm$^{-3}$). By using smaller $\kappa$, about 40\% higher $ZT$ can be obtained, as shown in Fig. \ref{fig2} (d), with a peak $ZT$ value up to 0.7 at 1400 K and 0.45 at 1000 K with similar doping level ($\sim$3-4$\times10^{20}$ cm$^{-3}$). Though a little higher in the latter case, our calculated $ZT$ with both experimental range $\kappa$ agree reasonably well with existing experiments ($\sim$0.2-0.4 at 900 K - 1000 K \cite{r16,r17,r18,r19,r38}). Complex isoenergy surfaces are favorable for electronic
performance as mentioned above. Therefore we studied the isoenergy surfaces of cubic STO especially at the peak $ZT$ values, for instance at 1400 K and 600 K with doping concentration of 5$\times10^{20}$ cm$^{-3}$ and 3$\times10^{20}$ cm$^{-3}$ as shown in Fig. \ref{fig2} (c). However we found the Fermi energy is about 0.12 eV (for 1400 K) below and 0.002 eV (600 K) above CBM at corresponding carrier concentrations due to the heavy onset feature at the CBM. Thus we present the isoenergy surfaces at 0.05 eV and 0.15 eV above CBM. As seen in Fig. \ref{fig3}, the surfaces show complex shapes due to the three lowest $t_{2g}$ orbitals. Specifically, the surfaces in the 0.15 eV case consist of intersecting cylinders joined at the center around $\Gamma$ and two small sections inside the center, as was discussed previously for STO and other $d^{0}$cubic perovskites \cite{r41}. This low-dimensional-like intersecting cylinders in higher carrier concentration (right panel of Fig. \ref{fig3}) that originated from the $t_{2g}$ orbitals is beneficial for achieving high $ZT$. Thus high doping concentration will be advantageous in achieving high thermoelectric performance for cubic STO.

\section{\label{sec:level4}SUMMARY AND CONCLUSIONS}
In summary, we have investigated the electronic and thermoelectric properties
of SrTiO$_{3}$. Using the combination of
constant scattering time approximation Boltzmann theory
based on first-principles electronic structure and existing experiments,
we have predicted the thermoelectric performance of STO,
including the Figure of merit, $ZT$.
Importantly, we find high $ZT$ value of 0.7 at high temperature (1400 K).
This confirms the potential of high temperature thermoelectric
properties of STO.
We find $ZT$ can be further improved relative to currently
reported experimental
values via carrier concentration optimization.

\begin{acknowledgments}
We are grateful for helpful discussions with Susanne Stemmer.
This work was supported by the Department of Energy through the S3TEC
Energy Frontier Research Center award \# DE-SC0001299/DE-FG02-09ER46577.
\end{acknowledgments}

\bibliography{SrTiO3}

\begin{thebibliography}{72}
\expandafter\ifx\csname natexlab\endcsname\relax\def\natexlab#1{#1}\fi
\expandafter\ifx\csname bibnamefont\endcsname\relax
  \def\bibnamefont#1{#1}\fi
\expandafter\ifx\csname bibfnamefont\endcsname\relax
  \def\bibfnamefont#1{#1}\fi
\expandafter\ifx\csname citenamefont\endcsname\relax
  \def\citenamefont#1{#1}\fi
\expandafter\ifx\csname url\endcsname\relax
  \def\url#1{\texttt{#1}}\fi
\expandafter\ifx\csname urlprefix\endcsname\relax\def\urlprefix{URL }\fi
\providecommand{\bibinfo}[2]{#2}
\providecommand{\eprint}[2][]{\url{#2}}

\bibitem[{\citenamefont{Wood}(1988)}]{r0}
\bibinfo{author}{\bibfnamefont{C.}~\bibnamefont{Wood}}, \bibinfo{journal}{Rep.
  Prog. Phys.} \textbf{\bibinfo{volume}{51}}, \bibinfo{pages}{459}
  (\bibinfo{year}{1988}).

\bibitem[{\citenamefont{Yang and Caillat}(2006)}]{r1}
\bibinfo{author}{\bibfnamefont{J.}~\bibnamefont{Yang}} \bibnamefont{and}
  \bibinfo{author}{\bibfnamefont{T.}~\bibnamefont{Caillat}},
  \bibinfo{journal}{MRS Bull.} \textbf{\bibinfo{volume}{31}},
  \bibinfo{pages}{224} (\bibinfo{year}{2006}).

\bibitem[{\citenamefont{Snyder and Toberer}(2008)}]{r2}
\bibinfo{author}{\bibfnamefont{G.~J.} \bibnamefont{Snyder}} \bibnamefont{and}
  \bibinfo{author}{\bibfnamefont{E.~S.} \bibnamefont{Toberer}},
  \bibinfo{journal}{Nat. Mater.} \textbf{\bibinfo{volume}{7}},
  \bibinfo{pages}{105} (\bibinfo{year}{2008}).

\bibitem[{\citenamefont{Matsubara et~al.}(2001)\citenamefont{Matsubara,
  Funahashi, Takeuchi, Sodeoka, Shimizu, and Ueno}}]{matsubara}
\bibinfo{author}{\bibfnamefont{I.}~\bibnamefont{Matsubara}},
  \bibinfo{author}{\bibfnamefont{R.}~\bibnamefont{Funahashi}},
  \bibinfo{author}{\bibfnamefont{T.}~\bibnamefont{Takeuchi}},
  \bibinfo{author}{\bibfnamefont{S.}~\bibnamefont{Sodeoka}},
  \bibinfo{author}{\bibfnamefont{T.}~\bibnamefont{Shimizu}}, \bibnamefont{and}
  \bibinfo{author}{\bibfnamefont{K.}~\bibnamefont{Ueno}},
  \bibinfo{journal}{Appl. Phys. Lett.} \textbf{\bibinfo{volume}{78}},
  \bibinfo{pages}{3627} (\bibinfo{year}{2001}).

\bibitem[{\citenamefont{Tsubota et~al.}(1997)\citenamefont{Tsubota, Ohtaki,
  Eguchi, and Arai}}]{tsubota}
\bibinfo{author}{\bibfnamefont{T.}~\bibnamefont{Tsubota}},
  \bibinfo{author}{\bibfnamefont{M.}~\bibnamefont{Ohtaki}},
  \bibinfo{author}{\bibfnamefont{K.}~\bibnamefont{Eguchi}}, \bibnamefont{and}
  \bibinfo{author}{\bibfnamefont{H.}~\bibnamefont{Arai}}, \bibinfo{journal}{J.
  Mater. Chem.} \textbf{\bibinfo{volume}{7}}, \bibinfo{pages}{85}
  (\bibinfo{year}{1997}).

\bibitem[{\citenamefont{He et~al.}(2011)\citenamefont{He, Liu, and
  Funahashi}}]{he}
\bibinfo{author}{\bibfnamefont{J.}~\bibnamefont{He}},
  \bibinfo{author}{\bibfnamefont{Y.}~\bibnamefont{Liu}}, \bibnamefont{and}
  \bibinfo{author}{\bibfnamefont{R.}~\bibnamefont{Funahashi}},
  \bibinfo{journal}{J. Mater. Res.} \textbf{\bibinfo{volume}{26}},
  \bibinfo{pages}{1762} (\bibinfo{year}{2011}).

\bibitem[{\citenamefont{Saucke et~al.}(2015)\citenamefont{Saucke, Populoh,
  Thiel, Xie, Funahashi, and Weidenkaff}}]{saucke}
\bibinfo{author}{\bibfnamefont{G.}~\bibnamefont{Saucke}},
  \bibinfo{author}{\bibfnamefont{S.}~\bibnamefont{Populoh}},
  \bibinfo{author}{\bibfnamefont{P.}~\bibnamefont{Thiel}},
  \bibinfo{author}{\bibfnamefont{W.~J.} \bibnamefont{Xie}},
  \bibinfo{author}{\bibfnamefont{R.}~\bibnamefont{Funahashi}},
  \bibnamefont{and}
  \bibinfo{author}{\bibfnamefont{A.}~\bibnamefont{Weidenkaff}},
  \bibinfo{journal}{J. Appl. Phys.} \textbf{\bibinfo{volume}{118}},
  \bibinfo{pages}{035106} (\bibinfo{year}{2015}).

\bibitem[{\citenamefont{Yamashita et~al.}(2003)\citenamefont{Yamashita,
  Tomiyoshi, and Makita}}]{r3}
\bibinfo{author}{\bibfnamefont{O.}~\bibnamefont{Yamashita}},
  \bibinfo{author}{\bibfnamefont{S.}~\bibnamefont{Tomiyoshi}},
  \bibnamefont{and} \bibinfo{author}{\bibfnamefont{K.}~\bibnamefont{Makita}},
  \bibinfo{journal}{J. Appl. Phys.} \textbf{\bibinfo{volume}{93}},
  \bibinfo{pages}{368} (\bibinfo{year}{2003}).

\bibitem[{\citenamefont{Shi et~al.}(2011)\citenamefont{Shi, Yang, Salvador,
  Chi, Cho, Wang, Bai, Yang, Zhang, and Chen}}]{r4}
\bibinfo{author}{\bibfnamefont{X.}~\bibnamefont{Shi}},
  \bibinfo{author}{\bibfnamefont{J.}~\bibnamefont{Yang}},
  \bibinfo{author}{\bibfnamefont{J.~R.} \bibnamefont{Salvador}},
  \bibinfo{author}{\bibfnamefont{M.}~\bibnamefont{Chi}},
  \bibinfo{author}{\bibfnamefont{J.~Y.} \bibnamefont{Cho}},
  \bibinfo{author}{\bibfnamefont{H.}~\bibnamefont{Wang}},
  \bibinfo{author}{\bibfnamefont{S.}~\bibnamefont{Bai}},
  \bibinfo{author}{\bibfnamefont{J.}~\bibnamefont{Yang}},
  \bibinfo{author}{\bibfnamefont{W.}~\bibnamefont{Zhang}}, \bibnamefont{and}
  \bibinfo{author}{\bibfnamefont{L.}~\bibnamefont{Chen}}, \bibinfo{journal}{J.
  Am. Chem. Soc.} \textbf{\bibinfo{volume}{133}}, \bibinfo{pages}{7837}
  (\bibinfo{year}{2011}).

\bibitem[{\citenamefont{Nolas et~al.}(2000)\citenamefont{Nolas, Kaeser,
  Littleton, and Tritt}}]{r5}
\bibinfo{author}{\bibfnamefont{G.~S.} \bibnamefont{Nolas}},
  \bibinfo{author}{\bibfnamefont{M.}~\bibnamefont{Kaeser}},
  \bibinfo{author}{\bibfnamefont{R.~T.} \bibnamefont{Littleton}},
  \bibnamefont{and} \bibinfo{author}{\bibfnamefont{T.~M.} \bibnamefont{Tritt}},
  \bibinfo{journal}{Appl. Phys. Lett.} \textbf{\bibinfo{volume}{77}},
  \bibinfo{pages}{1855} (\bibinfo{year}{2000}).

\bibitem[{\citenamefont{Terasaki et~al.}(1997)\citenamefont{Terasaki, Sasago,
  and Uchinokura}}]{r6}
\bibinfo{author}{\bibfnamefont{I.}~\bibnamefont{Terasaki}},
  \bibinfo{author}{\bibfnamefont{Y.}~\bibnamefont{Sasago}}, \bibnamefont{and}
  \bibinfo{author}{\bibfnamefont{K.}~\bibnamefont{Uchinokura}},
  \bibinfo{journal}{Phys. Rev. B} \textbf{\bibinfo{volume}{56}},
  \bibinfo{pages}{R12685} (\bibinfo{year}{1997}).

\bibitem[{\citenamefont{Koumoto et~al.}(2006)\citenamefont{Koumoto, Terasaki,
  and Funahashi}}]{r7}
\bibinfo{author}{\bibfnamefont{K.}~\bibnamefont{Koumoto}},
  \bibinfo{author}{\bibfnamefont{I.}~\bibnamefont{Terasaki}}, \bibnamefont{and}
  \bibinfo{author}{\bibfnamefont{R.}~\bibnamefont{Funahashi}},
  \bibinfo{journal}{MRS Bulletin} \textbf{\bibinfo{volume}{31}},
  \bibinfo{pages}{206} (\bibinfo{year}{2006}).

\bibitem[{\citenamefont{Fujita et~al.}(2001)\citenamefont{Fujita, Mochida, and
  Nakamura}}]{r8}
\bibinfo{author}{\bibfnamefont{K.}~\bibnamefont{Fujita}},
  \bibinfo{author}{\bibfnamefont{T.}~\bibnamefont{Mochida}}, \bibnamefont{and}
  \bibinfo{author}{\bibfnamefont{K.}~\bibnamefont{Nakamura}}, in
  \emph{\bibinfo{booktitle}{Thermoelectrics, 2001. Proceedings ICT 2001. XX
  International Conference on}} (\bibinfo{year}{2001}), pp.
  \bibinfo{pages}{168--171}.

\bibitem[{\citenamefont{Shikano and Funahashi}(2003)}]{r9}
\bibinfo{author}{\bibfnamefont{M.}~\bibnamefont{Shikano}} \bibnamefont{and}
  \bibinfo{author}{\bibfnamefont{R.}~\bibnamefont{Funahashi}},
  \bibinfo{journal}{Appl. Phys. Lett.} \textbf{\bibinfo{volume}{82}},
  \bibinfo{pages}{1851} (\bibinfo{year}{2003}).

\bibitem[{\citenamefont{Ong et~al.}(2011)\citenamefont{Ong, Singh, and
  Wu}}]{r10}
\bibinfo{author}{\bibfnamefont{K.~P.} \bibnamefont{Ong}},
  \bibinfo{author}{\bibfnamefont{D.~J.} \bibnamefont{Singh}}, \bibnamefont{and}
  \bibinfo{author}{\bibfnamefont{P.}~\bibnamefont{Wu}}, \bibinfo{journal}{Phys.
  Rev. B} \textbf{\bibinfo{volume}{83}}, \bibinfo{pages}{115110}
  (\bibinfo{year}{2011}).

\bibitem[{\citenamefont{Ohtaki et~al.}(2009)\citenamefont{Ohtaki, Araki, and
  Yamamoto}}]{r11}
\bibinfo{author}{\bibfnamefont{M.}~\bibnamefont{Ohtaki}},
  \bibinfo{author}{\bibfnamefont{K.}~\bibnamefont{Araki}}, \bibnamefont{and}
  \bibinfo{author}{\bibfnamefont{K.}~\bibnamefont{Yamamoto}},
  \bibinfo{journal}{J. Electron. Mater.} \textbf{\bibinfo{volume}{38}},
  \bibinfo{pages}{1234} (\bibinfo{year}{2009}).

\bibitem[{\citenamefont{Jood et~al.}(2011)\citenamefont{Jood, Mehta, Zhang,
  Peleckis, Wang, Siegel, Borca-Tasciuc, Dou, and Ramanath}}]{r12}
\bibinfo{author}{\bibfnamefont{P.}~\bibnamefont{Jood}},
  \bibinfo{author}{\bibfnamefont{R.~J.} \bibnamefont{Mehta}},
  \bibinfo{author}{\bibfnamefont{Y.}~\bibnamefont{Zhang}},
  \bibinfo{author}{\bibfnamefont{G.}~\bibnamefont{Peleckis}},
  \bibinfo{author}{\bibfnamefont{X.}~\bibnamefont{Wang}},
  \bibinfo{author}{\bibfnamefont{R.~W.} \bibnamefont{Siegel}},
  \bibinfo{author}{\bibfnamefont{T.}~\bibnamefont{Borca-Tasciuc}},
  \bibinfo{author}{\bibfnamefont{S.~X.} \bibnamefont{Dou}}, \bibnamefont{and}
  \bibinfo{author}{\bibfnamefont{G.}~\bibnamefont{Ramanath}},
  \bibinfo{journal}{Nano Lett.} \textbf{\bibinfo{volume}{11}},
  \bibinfo{pages}{4337} (\bibinfo{year}{2011}).

\bibitem[{\citenamefont{Jood et~al.}(2014)\citenamefont{Jood, Mehta, Zhang,
  Borca-Tasciuc, Dou, Singh, and Ramanath}}]{r13}
\bibinfo{author}{\bibfnamefont{P.}~\bibnamefont{Jood}},
  \bibinfo{author}{\bibfnamefont{R.~J.} \bibnamefont{Mehta}},
  \bibinfo{author}{\bibfnamefont{Y.}~\bibnamefont{Zhang}},
  \bibinfo{author}{\bibfnamefont{T.}~\bibnamefont{Borca-Tasciuc}},
  \bibinfo{author}{\bibfnamefont{S.~X.} \bibnamefont{Dou}},
  \bibinfo{author}{\bibfnamefont{D.~J.} \bibnamefont{Singh}}, \bibnamefont{and}
  \bibinfo{author}{\bibfnamefont{G.}~\bibnamefont{Ramanath}},
  \bibinfo{journal}{RSC Adv.} \textbf{\bibinfo{volume}{4}},
  \bibinfo{pages}{6363} (\bibinfo{year}{2014}).

\bibitem[{\citenamefont{Okuda et~al.}(2001)\citenamefont{Okuda, Nakanishi,
  Miyasaka, and Tokura}}]{r14}
\bibinfo{author}{\bibfnamefont{T.}~\bibnamefont{Okuda}},
  \bibinfo{author}{\bibfnamefont{K.}~\bibnamefont{Nakanishi}},
  \bibinfo{author}{\bibfnamefont{S.}~\bibnamefont{Miyasaka}}, \bibnamefont{and}
  \bibinfo{author}{\bibfnamefont{Y.}~\bibnamefont{Tokura}},
  \bibinfo{journal}{Phys. Rev. B} \textbf{\bibinfo{volume}{63}},
  \bibinfo{pages}{113104} (\bibinfo{year}{2001}).

\bibitem[{\citenamefont{Muta et~al.}(2005)\citenamefont{Muta, Kurosaki, and
  Yamanaka}}]{r15}
\bibinfo{author}{\bibfnamefont{H.}~\bibnamefont{Muta}},
  \bibinfo{author}{\bibfnamefont{K.}~\bibnamefont{Kurosaki}}, \bibnamefont{and}
  \bibinfo{author}{\bibfnamefont{S.}~\bibnamefont{Yamanaka}},
  \bibinfo{journal}{J. Alloys Compd} \textbf{\bibinfo{volume}{392}},
  \bibinfo{pages}{306 } (\bibinfo{year}{2005}).

\bibitem[{\citenamefont{Ohta et~al.}(2005)\citenamefont{Ohta, Nomura, Ohta, and
  Koumoto}}]{r16}
\bibinfo{author}{\bibfnamefont{S.}~\bibnamefont{Ohta}},
  \bibinfo{author}{\bibfnamefont{T.}~\bibnamefont{Nomura}},
  \bibinfo{author}{\bibfnamefont{H.}~\bibnamefont{Ohta}}, \bibnamefont{and}
  \bibinfo{author}{\bibfnamefont{K.}~\bibnamefont{Koumoto}},
  \bibinfo{journal}{J. Appl. Phys.} \textbf{\bibinfo{volume}{97}},
  \bibinfo{pages}{034106} (\bibinfo{year}{2005}).

\bibitem[{\citenamefont{Wang et~al.}(2011{\natexlab{a}})\citenamefont{Wang,
  Wang, Su, Liu, Sun, Peng, and Mei}}]{r17}
\bibinfo{author}{\bibfnamefont{H.~C.} \bibnamefont{Wang}},
  \bibinfo{author}{\bibfnamefont{C.~L.} \bibnamefont{Wang}},
  \bibinfo{author}{\bibfnamefont{W.~B.} \bibnamefont{Su}},
  \bibinfo{author}{\bibfnamefont{J.}~\bibnamefont{Liu}},
  \bibinfo{author}{\bibfnamefont{Y.}~\bibnamefont{Sun}},
  \bibinfo{author}{\bibfnamefont{H.}~\bibnamefont{Peng}}, \bibnamefont{and}
  \bibinfo{author}{\bibfnamefont{L.~M.} \bibnamefont{Mei}},
  \bibinfo{journal}{J. Am. Ceram. Soc.} \textbf{\bibinfo{volume}{94}},
  \bibinfo{pages}{838} (\bibinfo{year}{2011}{\natexlab{a}}).

\bibitem[{\citenamefont{Popuri et~al.}(2014)\citenamefont{Popuri, Scott,
  Downie, Hall, Suard, Decourt, Pollet, and Bos}}]{r18}
\bibinfo{author}{\bibfnamefont{S.~R.} \bibnamefont{Popuri}},
  \bibinfo{author}{\bibfnamefont{A.~J.~M.} \bibnamefont{Scott}},
  \bibinfo{author}{\bibfnamefont{R.~A.} \bibnamefont{Downie}},
  \bibinfo{author}{\bibfnamefont{M.~A.} \bibnamefont{Hall}},
  \bibinfo{author}{\bibfnamefont{E.}~\bibnamefont{Suard}},
  \bibinfo{author}{\bibfnamefont{R.}~\bibnamefont{Decourt}},
  \bibinfo{author}{\bibfnamefont{M.}~\bibnamefont{Pollet}}, \bibnamefont{and}
  \bibinfo{author}{\bibfnamefont{J.-W.~G.} \bibnamefont{Bos}},
  \bibinfo{journal}{RSC Adv.} \textbf{\bibinfo{volume}{4}},
  \bibinfo{pages}{33720} (\bibinfo{year}{2014}).

\bibitem[{\citenamefont{Lu et~al.}(2016)\citenamefont{Lu, Zhang, Lei, Sinclair,
  and Reaney}}]{r19}
\bibinfo{author}{\bibfnamefont{Z.}~\bibnamefont{Lu}},
  \bibinfo{author}{\bibfnamefont{H.}~\bibnamefont{Zhang}},
  \bibinfo{author}{\bibfnamefont{W.}~\bibnamefont{Lei}},
  \bibinfo{author}{\bibfnamefont{D.~C.} \bibnamefont{Sinclair}},
  \bibnamefont{and} \bibinfo{author}{\bibfnamefont{I.~M.}
  \bibnamefont{Reaney}}, \bibinfo{journal}{Chem. Mater.}
  \textbf{\bibinfo{volume}{28}}, \bibinfo{pages}{925} (\bibinfo{year}{2016}).

\bibitem[{\citenamefont{Parker and Singh}(2010)}]{r20}
\bibinfo{author}{\bibfnamefont{D.}~\bibnamefont{Parker}} \bibnamefont{and}
  \bibinfo{author}{\bibfnamefont{D.~J.} \bibnamefont{Singh}},
  \bibinfo{journal}{Phys. Rev. B} \textbf{\bibinfo{volume}{82}},
  \bibinfo{pages}{035204} (\bibinfo{year}{2010}).

\bibitem[{\citenamefont{Ziman}(1972)}]{ziman}
\bibinfo{author}{\bibfnamefont{J.~M.} \bibnamefont{Ziman}},
  \emph{\bibinfo{title}{{Principles of the Theory of Solids, 2nd Edition}}}
  (\bibinfo{publisher}{Cambridge University Press, Cambridge},
  \bibinfo{year}{1972}).

\bibitem[{\citenamefont{Madsen and Singh}(2006)}]{r21}
\bibinfo{author}{\bibfnamefont{G.}~\bibnamefont{Madsen}} \bibnamefont{and}
  \bibinfo{author}{\bibfnamefont{D.~J.} \bibnamefont{Singh}},
  \bibinfo{journal}{Computer Phys. Commun.} \textbf{\bibinfo{volume}{175}},
  \bibinfo{pages}{67} (\bibinfo{year}{2006}).

\bibitem[{\citenamefont{Zhang and Singh}(2009)}]{zhang}
\bibinfo{author}{\bibfnamefont{L.}~\bibnamefont{Zhang}} \bibnamefont{and}
  \bibinfo{author}{\bibfnamefont{D.~J.} \bibnamefont{Singh}},
  \bibinfo{journal}{Phys. Rev. B} \textbf{\bibinfo{volume}{80}},
  \bibinfo{pages}{075117} (\bibinfo{year}{2009}).

\bibitem[{\citenamefont{Yang et~al.}(2008)\citenamefont{Yang, Li, Wu, Zhang,
  Chen, and Yang}}]{yang}
\bibinfo{author}{\bibfnamefont{J.}~\bibnamefont{Yang}},
  \bibinfo{author}{\bibfnamefont{H.~M.} \bibnamefont{Li}},
  \bibinfo{author}{\bibfnamefont{T.}~\bibnamefont{Wu}},
  \bibinfo{author}{\bibfnamefont{W.~Q.} \bibnamefont{Zhang}},
  \bibinfo{author}{\bibfnamefont{L.~D.} \bibnamefont{Chen}}, \bibnamefont{and}
  \bibinfo{author}{\bibfnamefont{J.~H.} \bibnamefont{Yang}},
  \bibinfo{journal}{Adv. Funct. Mater.} \textbf{\bibinfo{volume}{18}},
  \bibinfo{pages}{2880} (\bibinfo{year}{2008}).

\bibitem[{\citenamefont{Parker and Singh}(2011)}]{r22}
\bibinfo{author}{\bibfnamefont{D.}~\bibnamefont{Parker}} \bibnamefont{and}
  \bibinfo{author}{\bibfnamefont{D.~J.} \bibnamefont{Singh}},
  \bibinfo{journal}{Phys. Rev. X} \textbf{\bibinfo{volume}{1}},
  \bibinfo{pages}{021005} (\bibinfo{year}{2011}).

\bibitem[{\citenamefont{Fei et~al.}(2014)\citenamefont{Fei, Faghaninia,
  Soklaski, Yan, Lo, and Yang}}]{fei}
\bibinfo{author}{\bibfnamefont{R.}~\bibnamefont{Fei}},
  \bibinfo{author}{\bibfnamefont{A.}~\bibnamefont{Faghaninia}},
  \bibinfo{author}{\bibfnamefont{R.}~\bibnamefont{Soklaski}},
  \bibinfo{author}{\bibfnamefont{J.~A.} \bibnamefont{Yan}},
  \bibinfo{author}{\bibfnamefont{C.}~\bibnamefont{Lo}}, \bibnamefont{and}
  \bibinfo{author}{\bibfnamefont{L.}~\bibnamefont{Yang}},
  \bibinfo{journal}{Nano Lett.} \textbf{\bibinfo{volume}{14}},
  \bibinfo{pages}{6393} (\bibinfo{year}{2014}).

\bibitem[{\citenamefont{Bjerg et~al.}(2011)\citenamefont{Bjerg, Madsen, and
  Iversen}}]{bjerg}
\bibinfo{author}{\bibfnamefont{L.}~\bibnamefont{Bjerg}},
  \bibinfo{author}{\bibfnamefont{G.~K.~H.} \bibnamefont{Madsen}},
  \bibnamefont{and} \bibinfo{author}{\bibfnamefont{B.~B.}
  \bibnamefont{Iversen}}, \bibinfo{journal}{Chem. Mater.}
  \textbf{\bibinfo{volume}{23}}, \bibinfo{pages}{3907} (\bibinfo{year}{2011}).

\bibitem[{\citenamefont{Singh and Kasinathan}(2007)}]{singh}
\bibinfo{author}{\bibfnamefont{D.~J.} \bibnamefont{Singh}} \bibnamefont{and}
  \bibinfo{author}{\bibfnamefont{D.}~\bibnamefont{Kasinathan}},
  \bibinfo{journal}{J. Electron. Mater.} \textbf{\bibinfo{volume}{36}},
  \bibinfo{pages}{736} (\bibinfo{year}{2007}).

\bibitem[{\citenamefont{Ong et~al.}(2010)\citenamefont{Ong, Singh, and
  Wu}}]{ong}
\bibinfo{author}{\bibfnamefont{K.~P.} \bibnamefont{Ong}},
  \bibinfo{author}{\bibfnamefont{D.~J.} \bibnamefont{Singh}}, \bibnamefont{and}
  \bibinfo{author}{\bibfnamefont{P.}~\bibnamefont{Wu}}, \bibinfo{journal}{Phys.
  Rev. Lett.} \textbf{\bibinfo{volume}{104}}, \bibinfo{pages}{176601}
  (\bibinfo{year}{2010}).

\bibitem[{\citenamefont{Moll et~al.}(2016)\citenamefont{Moll, Kushwaha, Nandi,
  Schmidt, and Mackenzie}}]{moll}
\bibinfo{author}{\bibfnamefont{P.~J.~W.} \bibnamefont{Moll}},
  \bibinfo{author}{\bibfnamefont{P.}~\bibnamefont{Kushwaha}},
  \bibinfo{author}{\bibfnamefont{N.}~\bibnamefont{Nandi}},
  \bibinfo{author}{\bibfnamefont{B.}~\bibnamefont{Schmidt}}, \bibnamefont{and}
  \bibinfo{author}{\bibfnamefont{A.~P.} \bibnamefont{Mackenzie}},
  \bibinfo{journal}{Science} \textbf{\bibinfo{volume}{351}},
  \bibinfo{pages}{1061} (\bibinfo{year}{2016}).

\bibitem[{\citenamefont{Kinaci et~al.}(2010)\citenamefont{Kinaci, Cevik, and
  Cagin}}]{kinaci}
\bibinfo{author}{\bibfnamefont{A.}~\bibnamefont{Kinaci}},
  \bibinfo{author}{\bibfnamefont{C.}~\bibnamefont{Cevik}}, \bibnamefont{and}
  \bibinfo{author}{\bibfnamefont{T.}~\bibnamefont{Cagin}},
  \bibinfo{journal}{Phys. Rev. B} \textbf{\bibinfo{volume}{82}},
  \bibinfo{pages}{155114} (\bibinfo{year}{2010}).

\bibitem[{\citenamefont{Madsen}(2006)}]{madsen}
\bibinfo{author}{\bibfnamefont{G.~K.~H.} \bibnamefont{Madsen}},
  \bibinfo{journal}{J. Am. Chem. Soc.} \textbf{\bibinfo{volume}{128}},
  \bibinfo{pages}{12140} (\bibinfo{year}{2006}).

\bibitem[{\citenamefont{Wang et~al.}(2011{\natexlab{b}})\citenamefont{Wang,
  Wang, Setyawan, Mingo, and Curtarolo}}]{wang}
\bibinfo{author}{\bibfnamefont{S.}~\bibnamefont{Wang}},
  \bibinfo{author}{\bibfnamefont{Z.}~\bibnamefont{Wang}},
  \bibinfo{author}{\bibfnamefont{W.}~\bibnamefont{Setyawan}},
  \bibinfo{author}{\bibfnamefont{N.}~\bibnamefont{Mingo}}, \bibnamefont{and}
  \bibinfo{author}{\bibfnamefont{S.}~\bibnamefont{Curtarolo}},
  \bibinfo{journal}{Phys. Rev. X} \textbf{\bibinfo{volume}{1}},
  \bibinfo{pages}{021012} (\bibinfo{year}{2011}{\natexlab{b}}).

\bibitem[{\citenamefont{Curtarolo et~al.}(2013)\citenamefont{Curtarolo, Hart,
  Nardelli, Mingo, Sanvito, and Levy}}]{curtarolo}
\bibinfo{author}{\bibfnamefont{S.}~\bibnamefont{Curtarolo}},
  \bibinfo{author}{\bibfnamefont{G.~L.~W.} \bibnamefont{Hart}},
  \bibinfo{author}{\bibfnamefont{M.~B.} \bibnamefont{Nardelli}},
  \bibinfo{author}{\bibfnamefont{N.}~\bibnamefont{Mingo}},
  \bibinfo{author}{\bibfnamefont{S.}~\bibnamefont{Sanvito}}, \bibnamefont{and}
  \bibinfo{author}{\bibfnamefont{O.}~\bibnamefont{Levy}},
  \bibinfo{journal}{Nature Materials} \textbf{\bibinfo{volume}{12}},
  \bibinfo{pages}{191} (\bibinfo{year}{2013}).

\bibitem[{\citenamefont{Du and Singh}(2010)}]{du}
\bibinfo{author}{\bibfnamefont{M.~H.} \bibnamefont{Du}} \bibnamefont{and}
  \bibinfo{author}{\bibfnamefont{D.~J.} \bibnamefont{Singh}},
  \bibinfo{journal}{Phys. Rev. B} \textbf{\bibinfo{volume}{81}},
  \bibinfo{pages}{144114} (\bibinfo{year}{2010}).

\bibitem[{\citenamefont{Himmetoglu et~al.}(2014)\citenamefont{Himmetoglu,
  Janotti, Peelaers, Alkauskas, and {Van de Walle}}}]{himmetoglu}
\bibinfo{author}{\bibfnamefont{B.}~\bibnamefont{Himmetoglu}},
  \bibinfo{author}{\bibfnamefont{A.}~\bibnamefont{Janotti}},
  \bibinfo{author}{\bibfnamefont{H.}~\bibnamefont{Peelaers}},
  \bibinfo{author}{\bibfnamefont{A.}~\bibnamefont{Alkauskas}},
  \bibnamefont{and} \bibinfo{author}{\bibfnamefont{C.~G.} \bibnamefont{{Van de
  Walle}}}, \bibinfo{journal}{Phys. Rev. B} \textbf{\bibinfo{volume}{90}},
  \bibinfo{pages}{241204} (\bibinfo{year}{2014}).

\bibitem[{\citenamefont{Komirenko et~al.}(2000)\citenamefont{Komirenko, Kim,
  Stroscio, and Dutta}}]{komirenko}
\bibinfo{author}{\bibfnamefont{S.~M.} \bibnamefont{Komirenko}},
  \bibinfo{author}{\bibfnamefont{K.~W.} \bibnamefont{Kim}},
  \bibinfo{author}{\bibfnamefont{M.~A.} \bibnamefont{Stroscio}},
  \bibnamefont{and} \bibinfo{author}{\bibfnamefont{M.}~\bibnamefont{Dutta}},
  \bibinfo{journal}{Phys. Rev. B} \textbf{\bibinfo{volume}{61}},
  \bibinfo{pages}{2034} (\bibinfo{year}{2000}).

\bibitem[{\citenamefont{Ravich et~al.}(1971{\natexlab{a}})\citenamefont{Ravich,
  Efimova, and Tamarchenko}}]{ravich1}
\bibinfo{author}{\bibfnamefont{Y.~I.} \bibnamefont{Ravich}},
  \bibinfo{author}{\bibfnamefont{B.~A.} \bibnamefont{Efimova}},
  \bibnamefont{and} \bibinfo{author}{\bibfnamefont{V.~I.}
  \bibnamefont{Tamarchenko}}, \bibinfo{journal}{Phys. Status Solidi B}
  \textbf{\bibinfo{volume}{43}}, \bibinfo{pages}{11}
  (\bibinfo{year}{1971}{\natexlab{a}}).

\bibitem[{\citenamefont{Ravich et~al.}(1971{\natexlab{b}})\citenamefont{Ravich,
  Efimova, and Tamarchenko}}]{ravich2}
\bibinfo{author}{\bibfnamefont{Y.~I.} \bibnamefont{Ravich}},
  \bibinfo{author}{\bibfnamefont{B.~A.} \bibnamefont{Efimova}},
  \bibnamefont{and} \bibinfo{author}{\bibfnamefont{V.~I.}
  \bibnamefont{Tamarchenko}}, \bibinfo{journal}{Phys. Status Solidi B}
  \textbf{\bibinfo{volume}{43}}, \bibinfo{pages}{453}
  (\bibinfo{year}{1971}{\natexlab{b}}).

\bibitem[{\citenamefont{Ravich}(1970)}]{ravich3}
\bibinfo{author}{\bibfnamefont{Y.~I.} \bibnamefont{Ravich}},
  \emph{\bibinfo{title}{Semiconducting Lead Chalcogenides}}
  (\bibinfo{publisher}{Springer, Berlin}, \bibinfo{year}{1970}).

\bibitem[{\citenamefont{Singh}(2010{\natexlab{a}})}]{singh-pbte}
\bibinfo{author}{\bibfnamefont{D.~J.} \bibnamefont{Singh}},
  \bibinfo{journal}{Phys. Rev. B} \textbf{\bibinfo{volume}{81}},
  \bibinfo{pages}{195217} (\bibinfo{year}{2010}{\natexlab{a}}).

\bibitem[{\citenamefont{Pei et~al.}(2011)\citenamefont{Pei, {LaLonde}, Iwanaga,
  and Snyder}}]{pei}
\bibinfo{author}{\bibfnamefont{Y.}~\bibnamefont{Pei}},
  \bibinfo{author}{\bibfnamefont{A.}~\bibnamefont{{LaLonde}}},
  \bibinfo{author}{\bibfnamefont{S.}~\bibnamefont{Iwanaga}}, \bibnamefont{and}
  \bibinfo{author}{\bibfnamefont{G.~J.} \bibnamefont{Snyder}},
  \bibinfo{journal}{Energy Environ. Sci.} \textbf{\bibinfo{volume}{4}},
  \bibinfo{pages}{2085} (\bibinfo{year}{2011}).

\bibitem[{\citenamefont{Singh and Nordstrom}(2006)}]{r25}
\bibinfo{author}{\bibfnamefont{D.~J.} \bibnamefont{Singh}} \bibnamefont{and}
  \bibinfo{author}{\bibfnamefont{L.}~\bibnamefont{Nordstrom}},
  \emph{\bibinfo{title}{Planewaves, Pseudopotentials and the LAPW Method, 2nd
  Edition}} (\bibinfo{publisher}{Springer, Berlin}, \bibinfo{year}{2006}).

\bibitem[{\citenamefont{Schwarz et~al.}(2002)\citenamefont{Schwarz, Blaha, and
  Madsen}}]{r26}
\bibinfo{author}{\bibfnamefont{K.}~\bibnamefont{Schwarz}},
  \bibinfo{author}{\bibfnamefont{P.}~\bibnamefont{Blaha}}, \bibnamefont{and}
  \bibinfo{author}{\bibfnamefont{G.}~\bibnamefont{Madsen}},
  \bibinfo{journal}{Computer Phys. Commun.} \textbf{\bibinfo{volume}{147}},
  \bibinfo{pages}{71} (\bibinfo{year}{2002}).

\bibitem[{\citenamefont{Cao et~al.}(2000)\citenamefont{Cao, Sozontov, and
  Zegenhagen}}]{r27}
\bibinfo{author}{\bibfnamefont{L.}~\bibnamefont{Cao}},
  \bibinfo{author}{\bibfnamefont{E.}~\bibnamefont{Sozontov}}, \bibnamefont{and}
  \bibinfo{author}{\bibfnamefont{J.}~\bibnamefont{Zegenhagen}},
  \bibinfo{journal}{physica status solidi (a)} \textbf{\bibinfo{volume}{181}},
  \bibinfo{pages}{387} (\bibinfo{year}{2000}).

\bibitem[{\citenamefont{Tran and Blaha}(2009)}]{r28}
\bibinfo{author}{\bibfnamefont{F.}~\bibnamefont{Tran}} \bibnamefont{and}
  \bibinfo{author}{\bibfnamefont{P.}~\bibnamefont{Blaha}},
  \bibinfo{journal}{Phys. Rev. Lett.} \textbf{\bibinfo{volume}{102}}
  (\bibinfo{year}{2009}).

\bibitem[{\citenamefont{Singh}(2010{\natexlab{b}})}]{r29}
\bibinfo{author}{\bibfnamefont{D.~J.} \bibnamefont{Singh}},
  \bibinfo{journal}{Phys. Rev. B} \textbf{\bibinfo{volume}{82}},
  \bibinfo{pages}{205102} (\bibinfo{year}{2010}{\natexlab{b}}).

\bibitem[{\citenamefont{Cardona}(1965)}]{cardona}
\bibinfo{author}{\bibfnamefont{M.}~\bibnamefont{Cardona}},
  \bibinfo{journal}{Phys. Rev.} \textbf{\bibinfo{volume}{140}},
  \bibinfo{pages}{651} (\bibinfo{year}{1965}).

\bibitem[{\citenamefont{Mattheiss}(1972)}]{r30}
\bibinfo{author}{\bibfnamefont{L.~F.} \bibnamefont{Mattheiss}},
  \bibinfo{journal}{Phys. Rev. B} \textbf{\bibinfo{volume}{6}},
  \bibinfo{pages}{4718} (\bibinfo{year}{1972}).

\bibitem[{\citenamefont{Baniecki et~al.}(2013)\citenamefont{Baniecki, Ishii,
  Aso, Kurihara, and Ricinschi}}]{r31}
\bibinfo{author}{\bibfnamefont{J.~D.} \bibnamefont{Baniecki}},
  \bibinfo{author}{\bibfnamefont{M.}~\bibnamefont{Ishii}},
  \bibinfo{author}{\bibfnamefont{H.}~\bibnamefont{Aso}},
  \bibinfo{author}{\bibfnamefont{K.}~\bibnamefont{Kurihara}}, \bibnamefont{and}
  \bibinfo{author}{\bibfnamefont{D.}~\bibnamefont{Ricinschi}},
  \bibinfo{journal}{J. Appl. Phys.} \textbf{\bibinfo{volume}{113}},
  \bibinfo{pages}{013701} (\bibinfo{year}{2013}).

\bibitem[{\citenamefont{Uwe et~al.}(1985)\citenamefont{Uwe, Sakudo, and
  Yamaguchi}}]{r32}
\bibinfo{author}{\bibfnamefont{H.}~\bibnamefont{Uwe}},
  \bibinfo{author}{\bibfnamefont{T.}~\bibnamefont{Sakudo}}, \bibnamefont{and}
  \bibinfo{author}{\bibfnamefont{H.}~\bibnamefont{Yamaguchi}},
  \bibinfo{journal}{Jpn. J. Appl. Phys.} \textbf{\bibinfo{volume}{24}},
  \bibinfo{pages}{519} (\bibinfo{year}{1985}).

\bibitem[{\citenamefont{Frederikse et~al.}(1964)\citenamefont{Frederikse,
  Thurber, and Hosler}}]{r33}
\bibinfo{author}{\bibfnamefont{H.~P.~R.} \bibnamefont{Frederikse}},
  \bibinfo{author}{\bibfnamefont{W.~R.} \bibnamefont{Thurber}},
  \bibnamefont{and} \bibinfo{author}{\bibfnamefont{W.~R.}
  \bibnamefont{Hosler}}, \bibinfo{journal}{Phys. Rev.}
  \textbf{\bibinfo{volume}{134}}, \bibinfo{pages}{A442} (\bibinfo{year}{1964}).

\bibitem[{\citenamefont{Ohta et~al.}(2008)\citenamefont{Ohta, Sugiura, and
  Koumoto}}]{r34}
\bibinfo{author}{\bibfnamefont{H.}~\bibnamefont{Ohta}},
  \bibinfo{author}{\bibfnamefont{K.}~\bibnamefont{Sugiura}}, \bibnamefont{and}
  \bibinfo{author}{\bibfnamefont{K.}~\bibnamefont{Koumoto}},
  \bibinfo{journal}{Inorg. Chem.} \textbf{\bibinfo{volume}{47}},
  \bibinfo{pages}{8429} (\bibinfo{year}{2008}).

\bibitem[{\citenamefont{Xing et~al.}(2016)\citenamefont{Xing, Sun, Ong, Fan,
  Zheng, and Singh}}]{r41}
\bibinfo{author}{\bibfnamefont{G.}~\bibnamefont{Xing}},
  \bibinfo{author}{\bibfnamefont{J.}~\bibnamefont{Sun}},
  \bibinfo{author}{\bibfnamefont{K.~P.} \bibnamefont{Ong}},
  \bibinfo{author}{\bibfnamefont{X.}~\bibnamefont{Fan}},
  \bibinfo{author}{\bibfnamefont{W.}~\bibnamefont{Zheng}}, \bibnamefont{and}
  \bibinfo{author}{\bibfnamefont{D.~J.} \bibnamefont{Singh}},
  \bibinfo{journal}{APL Mater.} \textbf{\bibinfo{volume}{4}},
  \bibinfo{pages}{053201} (\bibinfo{year}{2016}).

\bibitem[{\citenamefont{Kuroki and Arita}(2007)}]{r42}
\bibinfo{author}{\bibfnamefont{K.}~\bibnamefont{Kuroki}} \bibnamefont{and}
  \bibinfo{author}{\bibfnamefont{R.}~\bibnamefont{Arita}}, \bibinfo{journal}{J.
  Phys. Soc. Jpn} \textbf{\bibinfo{volume}{76}}, \bibinfo{pages}{083707}
  (\bibinfo{year}{2007}).

\bibitem[{\citenamefont{Chen et~al.}(2013)\citenamefont{Chen, Parker, and
  Singh}}]{r43}
\bibinfo{author}{\bibfnamefont{X.}~\bibnamefont{Chen}},
  \bibinfo{author}{\bibfnamefont{D.}~\bibnamefont{Parker}}, \bibnamefont{and}
  \bibinfo{author}{\bibfnamefont{D.~J.} \bibnamefont{Singh}},
  \bibinfo{journal}{Sci. Rep.} \textbf{\bibinfo{volume}{3}},
  \bibinfo{pages}{3168} (\bibinfo{year}{2013}).

\bibitem[{\citenamefont{Usui et~al.}(2013)\citenamefont{Usui, Suzuki, Kuroki,
  Nakano, Kudo, and Nohara}}]{r44}
\bibinfo{author}{\bibfnamefont{H.}~\bibnamefont{Usui}},
  \bibinfo{author}{\bibfnamefont{K.}~\bibnamefont{Suzuki}},
  \bibinfo{author}{\bibfnamefont{K.}~\bibnamefont{Kuroki}},
  \bibinfo{author}{\bibfnamefont{S.}~\bibnamefont{Nakano}},
  \bibinfo{author}{\bibfnamefont{K.}~\bibnamefont{Kudo}}, \bibnamefont{and}
  \bibinfo{author}{\bibfnamefont{M.}~\bibnamefont{Nohara}},
  \bibinfo{journal}{Phys. Rev. B} \textbf{\bibinfo{volume}{88}},
  \bibinfo{pages}{075140} (\bibinfo{year}{2013}).

\bibitem[{\citenamefont{Usui et~al.}(2010)\citenamefont{Usui, Shibata, and
  Kuroki}}]{r45}
\bibinfo{author}{\bibfnamefont{H.}~\bibnamefont{Usui}},
  \bibinfo{author}{\bibfnamefont{S.}~\bibnamefont{Shibata}}, \bibnamefont{and}
  \bibinfo{author}{\bibfnamefont{K.}~\bibnamefont{Kuroki}},
  \bibinfo{journal}{Phys. Rev. B} \textbf{\bibinfo{volume}{81}},
  \bibinfo{pages}{205121} (\bibinfo{year}{2010}).

\bibitem[{\citenamefont{Shirai and Yamanaka}(2013)}]{r46}
\bibinfo{author}{\bibfnamefont{K.}~\bibnamefont{Shirai}} \bibnamefont{and}
  \bibinfo{author}{\bibfnamefont{K.}~\bibnamefont{Yamanaka}},
  \bibinfo{journal}{J. Appl. Phys.} \textbf{\bibinfo{volume}{113}},
  \bibinfo{pages}{053705} (\bibinfo{year}{2013}).

\bibitem[{\citenamefont{Parker et~al.}(2013)\citenamefont{Parker, Chen, and
  Singh}}]{r47}
\bibinfo{author}{\bibfnamefont{D.}~\bibnamefont{Parker}},
  \bibinfo{author}{\bibfnamefont{X.}~\bibnamefont{Chen}}, \bibnamefont{and}
  \bibinfo{author}{\bibfnamefont{D.~J.} \bibnamefont{Singh}},
  \bibinfo{journal}{Phys. Rev. Lett.} \textbf{\bibinfo{volume}{110}},
  \bibinfo{pages}{146601} (\bibinfo{year}{2013}).

\bibitem[{\citenamefont{Mecholsky et~al.}(2014)\citenamefont{Mecholsky, Resca,
  Pegg, and Fornari}}]{r48}
\bibinfo{author}{\bibfnamefont{N.~A.} \bibnamefont{Mecholsky}},
  \bibinfo{author}{\bibfnamefont{L.}~\bibnamefont{Resca}},
  \bibinfo{author}{\bibfnamefont{I.~L.} \bibnamefont{Pegg}}, \bibnamefont{and}
  \bibinfo{author}{\bibfnamefont{M.}~\bibnamefont{Fornari}},
  \bibinfo{journal}{Phys. Rev. B} \textbf{\bibinfo{volume}{89}},
  \bibinfo{pages}{155131} (\bibinfo{year}{2014}).

\bibitem[{\citenamefont{Sun and Singh}(2016)}]{r35}
\bibinfo{author}{\bibfnamefont{J.}~\bibnamefont{Sun}} \bibnamefont{and}
  \bibinfo{author}{\bibfnamefont{D.~J.} \bibnamefont{Singh}},
  \bibinfo{journal}{Phys. Rev. Applied} \textbf{\bibinfo{volume}{5}},
  \bibinfo{pages}{024006} (\bibinfo{year}{2016}).

\bibitem[{\citenamefont{Cain et~al.}(2013)\citenamefont{Cain, Kajdos, and
  Stemmer}}]{r36}
\bibinfo{author}{\bibfnamefont{T.~A.} \bibnamefont{Cain}},
  \bibinfo{author}{\bibfnamefont{A.~P.} \bibnamefont{Kajdos}},
  \bibnamefont{and} \bibinfo{author}{\bibfnamefont{S.}~\bibnamefont{Stemmer}},
  \bibinfo{journal}{Appl. Phys. Lett.} \textbf{\bibinfo{volume}{102}},
  \bibinfo{pages}{182101} (\bibinfo{year}{2013}).

\bibitem[{\citenamefont{Yu et~al.}(2008)\citenamefont{Yu, Scullin, Huijben,
  Ramesh, and Majumdar}}]{r37}
\bibinfo{author}{\bibfnamefont{C.}~\bibnamefont{Yu}},
  \bibinfo{author}{\bibfnamefont{M.~L.} \bibnamefont{Scullin}},
  \bibinfo{author}{\bibfnamefont{M.}~\bibnamefont{Huijben}},
  \bibinfo{author}{\bibfnamefont{R.}~\bibnamefont{Ramesh}}, \bibnamefont{and}
  \bibinfo{author}{\bibfnamefont{A.}~\bibnamefont{Majumdar}},
  \bibinfo{journal}{Appl. Phys. Lett.} \textbf{\bibinfo{volume}{92}},
  \bibinfo{pages}{191911} (\bibinfo{year}{2008}).

\bibitem[{\citenamefont{Wang et~al.}(2013)\citenamefont{Wang, Chen, He,
  Norimatsu, Kusunoki, and Koumoto}}]{r38}
\bibinfo{author}{\bibfnamefont{N.}~\bibnamefont{Wang}},
  \bibinfo{author}{\bibfnamefont{H.}~\bibnamefont{Chen}},
  \bibinfo{author}{\bibfnamefont{H.}~\bibnamefont{He}},
  \bibinfo{author}{\bibfnamefont{W.}~\bibnamefont{Norimatsu}},
  \bibinfo{author}{\bibfnamefont{M.}~\bibnamefont{Kusunoki}}, \bibnamefont{and}
  \bibinfo{author}{\bibfnamefont{K.}~\bibnamefont{Koumoto}},
  \bibinfo{journal}{Sci. Rep.} \textbf{\bibinfo{volume}{3}},
  \bibinfo{pages}{3449} (\bibinfo{year}{2013}).

\bibitem[{\citenamefont{Oh et~al.}(2011)\citenamefont{Oh, Ravichandran, Liang,
  Siemons, Jalan, Brooks, Huijben, Schlom, Stemmer, Martin et~al.}}]{r39}
\bibinfo{author}{\bibfnamefont{D.-W.} \bibnamefont{Oh}},
  \bibinfo{author}{\bibfnamefont{J.}~\bibnamefont{Ravichandran}},
  \bibinfo{author}{\bibfnamefont{C.-W.} \bibnamefont{Liang}},
  \bibinfo{author}{\bibfnamefont{W.}~\bibnamefont{Siemons}},
  \bibinfo{author}{\bibfnamefont{B.}~\bibnamefont{Jalan}},
  \bibinfo{author}{\bibfnamefont{C.~M.} \bibnamefont{Brooks}},
  \bibinfo{author}{\bibfnamefont{M.}~\bibnamefont{Huijben}},
  \bibinfo{author}{\bibfnamefont{D.~G.} \bibnamefont{Schlom}},
  \bibinfo{author}{\bibfnamefont{S.}~\bibnamefont{Stemmer}},
  \bibinfo{author}{\bibfnamefont{L.~W.} \bibnamefont{Martin}},
  \bibnamefont{et~al.}, \bibinfo{journal}{Appl. Phys. Lett.}
  \textbf{\bibinfo{volume}{98}}, \bibinfo{pages}{221904}
  (\bibinfo{year}{2011}).

\bibitem[{\citenamefont{Muta et~al.}(2003)\citenamefont{Muta, Kurosaki, and
  Yamanaka}}]{r40}
\bibinfo{author}{\bibfnamefont{H.}~\bibnamefont{Muta}},
  \bibinfo{author}{\bibfnamefont{K.}~\bibnamefont{Kurosaki}}, \bibnamefont{and}
  \bibinfo{author}{\bibfnamefont{S.}~\bibnamefont{Yamanaka}},
  \bibinfo{journal}{J. Alloys Compd.} \textbf{\bibinfo{volume}{350}},
  \bibinfo{pages}{292 } (\bibinfo{year}{2003}).

\end{thebibliography}

\end{document}